\documentclass[useAMS,usenatbib]{mn2e}

\usepackage{txfonts,amssymb}
\usepackage{amsfonts}
\usepackage{mncite}
\usepackage{epsfig}

\newcommand{\de}{\mbox{d}}

\title[The fragmentation boundary
revisited]{The response of self-graviting protostellar discs to
slow reduction in cooling timescale: the fragmentation boundary
revisited}

\author[C. Clarke, E. Harper-Clark \&  G. Lodato] {C.J.  Clarke  $^1$,
 E.
Harper-Clark$^2$, G. Lodato $^3$
 \\
$^1$Institute of Astronomy, Madingley Road, Cambridge, CB3 0HA \\
$^2$ Department of Astronomy and Astrophysics, 50 St. George Street, Toronto, Ontario, Canada, M5S 3H4 \\
$^3$ Department of Physics and Astronomy, University of Leicester, University Road, Leicester, LE1 7RH \\}

\begin{document}
\maketitle

\begin{abstract}

 A number of previous studies of the fragmentation of
self-gravitating protostellar discs have involved suites of simulations
in which
radiative cooling is modeled in terms of a cooling timescale ($t_{\rm
cool}$)
which is parameterised
as a simple multiple ($\beta_{\rm cool}$) of the local dynamical
timescale.  Such studies have
delineated the `fragmentation boundary' in terms of a critical value
of $\beta_{\rm cool}$ ($\beta_{\rm crit}$) such that the disc fragments if
$\beta_{\rm cool} < \beta_{\rm crit}$. Such an approach however begs
the question of how in reality a disc could ever be assembled
in a state with $\beta_{\rm cool} < \beta_{\rm crit}$. Here we adopt the
more realistic approach of effecting a gradual reduction in
$\beta_{\rm cool}$, as might correspond to changes in thermal
regime due to  secular changes in the disc density profile.
We find that the effect of {\it gradually}   reducing
$\beta_{\rm cool}$ (on a timescale longer than $t_{\rm cool}$) is to
stabilise the disc against fragmentation, compared with models
in which $\beta_{\rm cool}$ is reduced rapidly (over
less than $t_{\rm cool}$). We therefore conclude that the
ability of a disc to remain in a self-regulated, self-gravitating
state (without fragmentation) is partly dependent on the
disc's thermal history, as well as its  current cooling rate.
Nevertheless,
the effect of a slow reduction in $t_{\rm cool}$ appears only to
lower the fragmentation boundary by about a factor two in
$t_{\rm cool}$ and thus only permits maximum  `$\alpha$' values
(which parameterise the efficiency of angular momentum transfer
in the disc) 
that are about a factor two higher than determined hitherto. Our
results therefore do not undermine the notion that there is
a fundamental upper limit to the heating rate that can
be delivered by gravitational instabilities before the disc is
subject to fragmentation. An important implication of this work, therefore,
is that self-gravitating discs can enter into the regime of fragmentation
via {\it secular} evolution and it is not necessary to invoke rapid (impulsive)
events to trigger fragmentation.
\end{abstract}


\section{Introduction}

  Following the seminal work of \citet{gammie01}, there has been
considerable progress in recent years in understanding the behaviour of
self-gravitating accretion dics (see \citealt{PPV_GI} and references therein).
A number of simulations \citep{gammie01,rice03a,LR04,LR05}
have demonstrated that if the thermodynamic
properties of the disc are evolved according to a thermal equation
(involving a cooling term parameterised in terms of a cooling
timescale, $t_{\rm cool}$), then the disc may be able to establish a
{\it self-gravitating, self-regulated} state. In this state, the
Toomre $Q$ parameter:

\begin{equation}
Q=\frac{c_{\rm s}\kappa}{\pi G\Sigma},
\label{eq:Q}
\end{equation}
(where $c_{\rm s}$ is the sound speed, $\kappa$ is the epicyclic
frequency (equal to the angular velocity $\Omega$ in a Keplerian disc)
and $\Sigma$ is the disc surface density) hovers at a value somewhat
greater than
unity over an extended
region of the disc. Whereas the state $Q=1$, corresponds  to a situation
of marginal stability against {\it axisymmetric}  perturbations, in the
self-regulated  state the disc is instead subject to a variety of
non-axisymmetric self-gravitating modes whose effect, through the
action of weak shocks, is to dissipate mechanical energy (i.e. kinetic
and potential energy of the accretion flow) as heat.
Thermal equilibrium is then attained through the balancing of such
heating by the prescribed radiative cooling: in essence, self-regulation
results when the  amplitude of these modes is able to self-adjust so
as to maintain thermodynamic equilibrium against
the relevant  energy loss processes.

The above studies have all found, however, that such self-regulation
is only possible in the case that the cooling timescale is not
too short: stability demands that $ \beta_{\rm cool} = t_{\rm cool}/\Omega^{-1}
$
exceeds a critical value which, for discs with adiabatic index
of $5/3$, is $\sim 7$ \citep{RLA05}. In the case of more rapid cooling, the disc
instead fragments.

  Such simulations however approach the `fragmentation boundary'
in a manner that is unlikely ever to apply to discs in reality.
In the simulations, the discs are set up without additional heating
mechanisms and are subject to cooling at some prescribed value of
$\beta_{\rm cool}$. Discs with $\beta_{\rm cool}< \beta_{\rm crit}$
then fragment on the
local cooling timescale (i.e. a few times the local dynamical
timescale), thus begging the question of how such unstable initial
conditions could ever have been set up in the first place.

 A more likely scenario for disc  fragmentation is that
the disc is instead set up in self-regulated, self-gravitating
state and then  conditions {\it gradually} change so that $\beta_{\rm cool}$
is lowered. (For example, continued infall of material onto
a disc or secular re-arrangement of material in the disc due to the
action of gravitational torques could alter the surface density profile
of the disc and allow it  to enter a new cooling regime with
lower $\beta_{\rm cool}$). It is not however clear that the fragmentation
boundary would be the same in the case that $\beta_{\rm cool}$ is gradually
reduced.

 In this paper we conduct a suite of idealised simulations in which
we explore whether the fragmentation boundary just depends on the
instantaneous value of $\beta_{\rm cool}$ (as has been assumed hitherto)
of whether the system `remembers' the history of how it
evolved to a point of given $\beta_{\rm cool}$. 
Such a  (`toy model') approach, is complementary to studies
(\citealt{boley06,mayer07,stamatellos07}, see also the analytical estimates by \citealt{rafikov05,rafikov07}) which attempt to achieve ever-increasing verisimilitude
via the incorporation of more realistic treatments of radiative
transfer. Here, instead, we make no claims that the simplified
cooling law (for example, the assumption that $\beta_{\rm cool}$ is
spatially uniform) actually corresponds to a situation encountered
in a real disc, because our aim is to isolate a particular physical
effect  (i.e. the timescale on which the fragmentation boundary is approached).
The computational expense of `realistic' simulations however prevents 
their use to study secular effects:  even in the case of the
present `toy' simulations, 
it is impracticable
to run simulations over the long timescales on which the 
$\Sigma$ profile changes due to gravitational torques or infall.
We can nevertheless assess the effect of relatively slow changes
in $\beta_{\rm cool}$ on the  fragmentation boundary
through imposing 
an {\it ad hoc} reduction in the value of
$\beta_{\rm cool}$ and can apply  this insight to  the secular evolution
of real discs.   

 In particular, we want to examine the
cause of
the fragmentation for $\beta_{\rm cool} < \beta_{\rm crit}$, that has been found in
previous simulations. Is this (i)  due to the disc's  inability  to
maintain
 - under any circumstances - a gravitational
heating rate that can match the imposed high cooling rate?
This is the hypothesis of \citet{LR05}, who identify the minimum
value of $\beta_{\rm cool}$ with a maximum value of the gravitationally
induced angular momentum transfer that can be delivered by
a disc without its fragmenting. They parameterise this state of
maximal angular momentum transfer in terms of the ratio of the
${r,\phi}$ component of the stress tensor to the thermal pressure,
i.e., by analogy
with the equivalent expression for a {\it viscous} disc, in terms
of a maximum in the well known viscous `$\alpha$' parameter
\citep{shakura73}. A critical value of $\beta_{\rm cool}$
of $\sim 7$ corresponds to a maximum $\alpha$ of $\sim 0.06$.

 Alternatively, (ii) does fragmentation instead reflect the disc's
inability to set up
the required high heating rate {\it on the short timescale}
($t_{\rm cool}$) on which the disc is cooling?
If this were the case, then with sufficiently
gradual approach to the regime of low $\beta_{\rm cool}$, the disc
could in principle deliver a value of $\alpha$ that exceeded the above limit
by a generous margin.

We can obviously distinguish
between these alternatives by investigating the case in which $\beta_{\rm cool}$
is reduced on a timescale $\tau$ that is longer than $t_{\rm cool}$, since
in
this case the disc temperature will fall via a sequence of
thermal equilibrium states (on timescale $\tau$), rather than dropping
on timescale $t_{\rm cool}$.
The aim of this investigation is thus to
see whether the disc is more resistant to fragmentation in the
regime that $\tau > t_{\rm cool}$. If it is {\it not}, then 
the manner in which the disc approaches the fragmentation
boundary is  unimportant. If, on the other hand, it is found
that rapid changes in cooling regime are required, then it may 
be necessary to invoke  impulsive events (such as
an external dynamical interaction) to trigger fragmentation.

In Section 2 we describe the numerical setup, discuss  our results
in Section 3 and in Section 4 we present some conclusions.

\section{Numerical setup}
\label{sec:setup}

\subsection{The SPH code}

Our three-dimensional numerical simulations are carried out using SPH,
a Lagrangian hydrodynamic scheme \citep{benz90,monaghan92}.
The general implementation is very similar to \citet{LR04}, \citet{LR05}
and \citet{RLA05}.
 The gas disc is modeled with 250,000 SPH particles
(500,000 in a run used as a convergence test) and the local fluid
properties are computed by suitably averaging over the neighbouring
particles. The disc is set in almost Keplerian rotation (allowing from slight departures from it to account for the effect of pressure forces and of the disc gravitational force) around a central point
mass onto which gas particles can accrete if they get closer than the
accretion radius, taken to be equal to 0.5 code units.

The gas disc can heat up due to $p\mbox{d}V$ work and artificial
viscosity. The ratio of specific heats is $\gamma=5/3$. Cooling is here
implemented in a simplified way, i.e. by parameterizing the cooling
rate in terms of a cooling timescale:

\begin{equation}
\left(\frac{\de u_{\rm i}}{\de t}\right)_{\rm cool}=
-\frac{u_{\rm i}}{t_{\rm cool}},
\end{equation}
where $u_{\rm i}$ is the internal energy of a particle and the cooling
timescale $t_{\rm cool}$ is assumed to be proportional to the dynamical
timescale, $t_{\rm cool}=\beta_{\rm cool}\Omega^{-1}$, where $\beta_{\rm cool}$
is varied according to a time-dependent prescription (see Section 2.3
below).

Artificial viscosity is introduced using the standard SPH formalism. The actual implementiation is very similar to the one used in \citet{RLA05}, that is we set the two relevant numerical parameters to $\alpha_{\rm SPH}=0.1$ and $\beta_{\rm SPH}=0.2$ and we have not included here (consistent with \citealt{RLA05}) the so-called Balsara switch \citep{balsara95} to reduce shear viscosity. 

\begin{table}
\begin{tabular}{c|c|c|c|c}

Simulation & $x$  & $\beta_{\rm hold}$  & $N$ & fragmentation  \\
\hline

F1               &  10.5        & ---    & 250K& yes  \\
F2              &  10.5         & 3      & 250K& yes \\
V                &  105-10.5 & 3      & 250K& yes  \\
S1              & 105           & 3      & 250K &no  \\
Sh              &  105          & 3      & 500K & yes\\
S2              &  105          & 2.75 & 250K& no\\
S3              &   105         & 2.62 & 250K&yes\\
VS1           &    314         &  3     & 250K & no\\
VS2           &    314         &  2.75& 250K& yes\\
\hline

\end{tabular}
\caption{\small Details of the various simulations discussed in this paper. The different columns indicate: the name of the run, the value of the parameter $x$ determining the speed of the reduction of the cooling time, the value of $\beta_{\rm hold}$ (if any) at which the cooling time was held fixed after reduction, the number of particles used in the run $N$ and whether fragmentation did occur or not. Simulation V was performed with an initially slow reduction of $\beta$ (with $x=105$), followed by a fast reduction (with $x=10.5$), so that it would reach $\beta=3$ with a fast reduction at the same time as simulation S1.}
\label{tab:table}
\end{table}

\subsection{Disc setup}

The main physical properties of the disc at the beginning of the
simulation are again similar to those of \citet{LR04,LR05}.
The disc surface density $\Sigma$ is initially
proportional to $R^{-1}$ (where $R$ is the cylindrical radius), while
the temperature is initially proportional to $R^{-1/2}$. Given our
simplified form of the cooling function, the computations described
here are essentially scale free and can be rescaled to different disc
sizes and masses. For reference, we will assume that the unit mass
(which is the mass of the central star) is $1M_{\odot}$ and that the
unit radius is $1AU$. In this units the disc extends from $R_{\rm
in}=0.25 AU$ to $R_{\rm out}=25 AU$.  The normalization of the surface
density is generally chosen such as to have a total disc mass of
$M_{\rm disc}=0.1M_{\odot}$, while the
temperature normalization is chosen so as to have a minimum value of
$Q=2$, which is attained at the outer edge of the disc.

Initially, the disc is evolved with constant $\beta_{\rm cool}=7.5$, this
value of $\beta_{\rm cool}$ being in the regime
where previous work \citet{gammie01}, \citet{RLA05}
has shown that
the disc does not fragment. The general features of this initial
evolution is described
in detail in \citet{LR04}.
The disc starts cooling down until the
vertical scale-length $H$ is reduced such that $H/R\approx M_{\rm
disc}/M_{\star}=0.1$. At this point the disc becomes Toomre unstable
and develops a spiral structure that heats up the disc and maintains it
close to marginal stability. We have evolved the disc with this value of
$\beta_{\rm cool}$ for $7.8$ outer disc orbits. At this stage it
is close to $Q=1$ over most
of the disc (i.e. over the radial range $R=3-23$ A.U. ). 

\subsection{Evolution of $\beta_{ \rm cool}$}

After evolution of the disc with $\beta_{\rm cool} = \beta_{\rm cool} (0)=
7.5$ for a cooling timescale, we effect a linear reduction of $\beta_{ \rm cool}$ on
a timescale $T$, i.e.

\begin{equation}
\beta_{\rm cool}(t) = \beta_{\rm cool}(0)\left(1-\frac {t}{T}\right)
\end{equation}
where we set $T=x \Omega^{-1}(R_{out})$.

Such a prescription implies that the timescale $\tau(t)$ on which
the local instantaneous value of $t_{\rm cool}$ (i.e. $t_{\rm cool} (R,t)$) drops
to zero is
\begin{equation}
\tau(t) =  \frac {\beta_{\rm cool}(t)} {|\dot \beta_{\rm cool}|}=\frac{x}{\beta_{\rm cool}(0)} \left(\frac{R}{R_{out}}\right)^{-1.5} t_{\rm cool}(R,t).
\end{equation}

We adopt  three values of $x$: $x=10.5$ (fast), $x=105$
(slow) and $x=314$ (very slow). In the fast case, $\tau(R,t) \sim t_{\rm
cool}(R,t)$ in the outer disc so $t_{\rm cool}$ is changing faster
than the disc can come into thermal equilibrium at that value of
$t_{\rm cool}$. In the slow case, $\tau(r,t) > t_{\rm cool}(r,t)$
so that the disc is everywhere able to come into thermal equilibrium at 
that $t_{\rm cool}$. This situation is even more amply satisfied in  the very slow
case.

For each value of $x$, we run the simulation until a fragment forms
(at $\beta_{\rm cool}
= \beta_{\rm frag}$). In those cases where the timing of fragmentation
suggests that the slow reduction of $\beta_{\rm cool}$ is acting so as to
stabilise the disc at lower $\beta_{\rm cool}$, we test this
hypothesis by turning off the reduction in $\beta_{\rm cool}$  when it
attains a value equal to
$\beta_{\rm hold} (> \beta_{\rm frag}$). We then
experiment with values of
$\beta_{\rm hold}$ in order to find the minimum value of
$\beta_{\rm hold}$ at which the disc does not fragment over the
duration of the numerical experiment. Table \ref{tab:table} and Fig. 1 summarize the main details of the various runs we have performed, where the `F' simulations are the `fast' ones, the `S' are the `slow' ones and the `VS' are the `very slow' ones (see discussion in Section 3 below). The simulation named V was performed with an initially slow reduction of $\beta$ (with $x=105$), followed by a fast reduction (with $x=10.5$), so that it would reach $\beta=3$ with a fast reduction at the same time as simulation S1. This was run as a control run to ensure that secular evolution did not affect our results (see below).

\begin{figure}
            \epsfig{figure=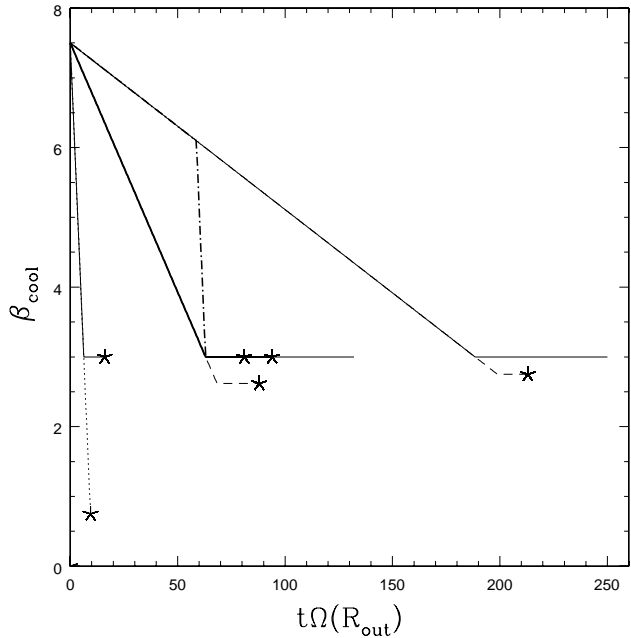,width=0.5\textwidth}
\caption{\small {Time evolution of the parameter $\beta_{\rm cool}$ in the various models. The three solid lines refer to cases where $\beta_{\rm cool}$ is first decreased and then held at $\beta_{\rm cool}=3$ (that is, simulations F2, S1 and VS1). The two dashed lines correspond to simulations S3 and VS2. The dotted line is simulation F1 and the thick solid line is the higher resolution run (Sh). Finally, the dot-dashed line is the simulation with variable rate of change of $\beta_{\rm cool}$ (V). 
An asterisk  at the end of the line indicates fragmentation at this time,
whereas runs without an asterisk are unfragmented at the end of  the simulation.}}
\end{figure}

\begin{figure*}
\centerline{\epsfig{figure=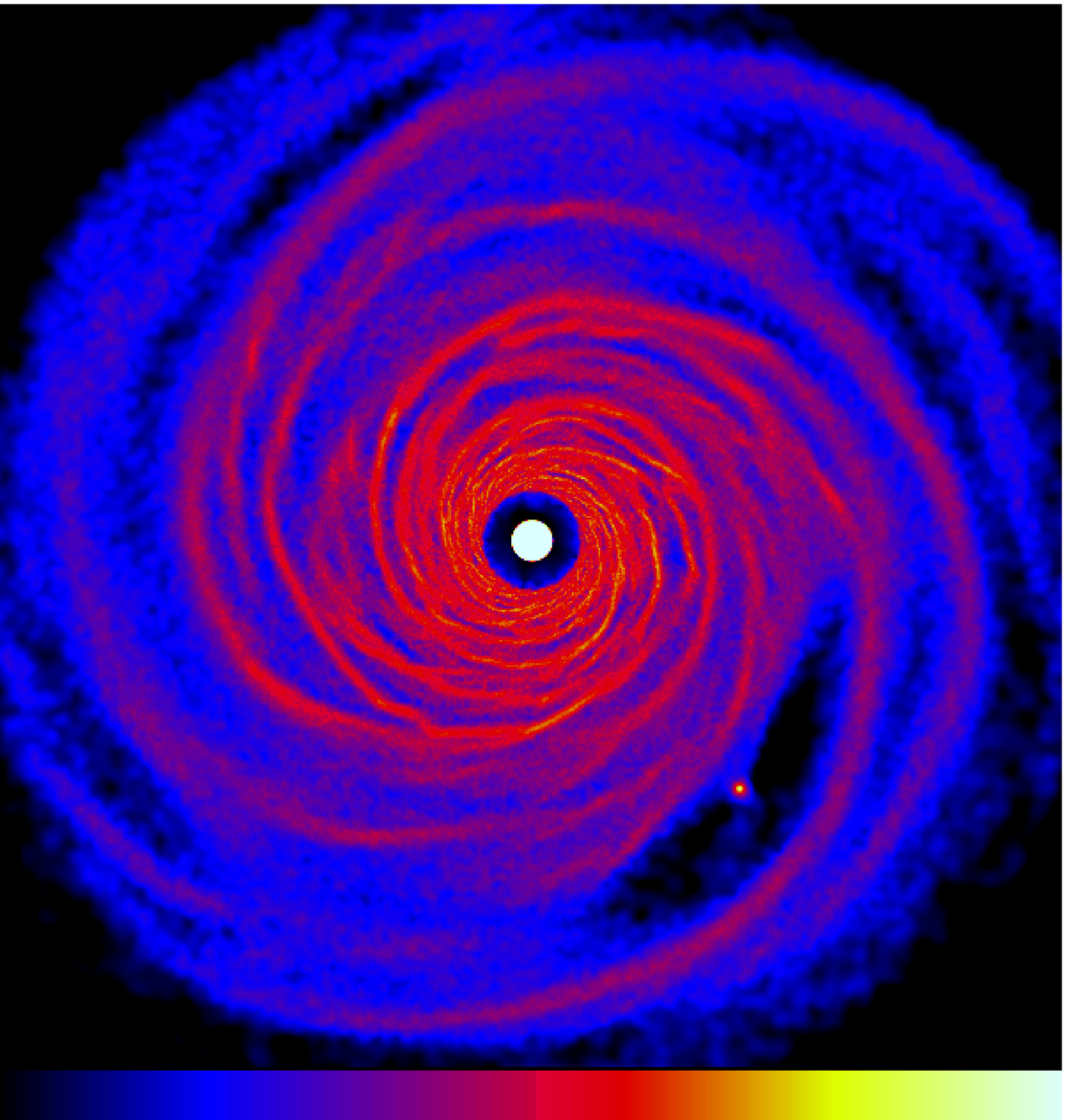,width=0.3\textwidth}
            \epsfig{figure=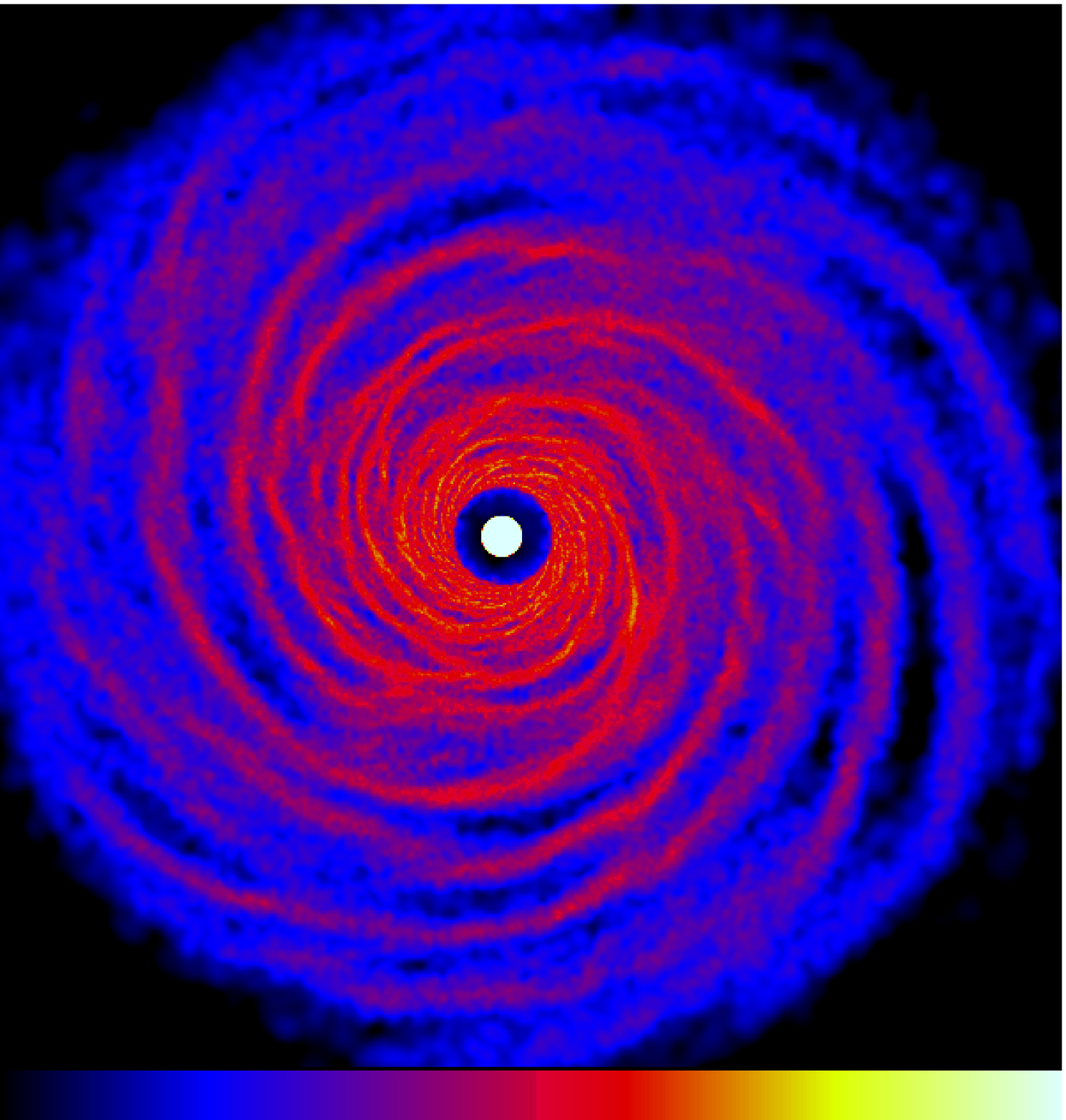,width=0.3\textwidth}
             \epsfig{figure=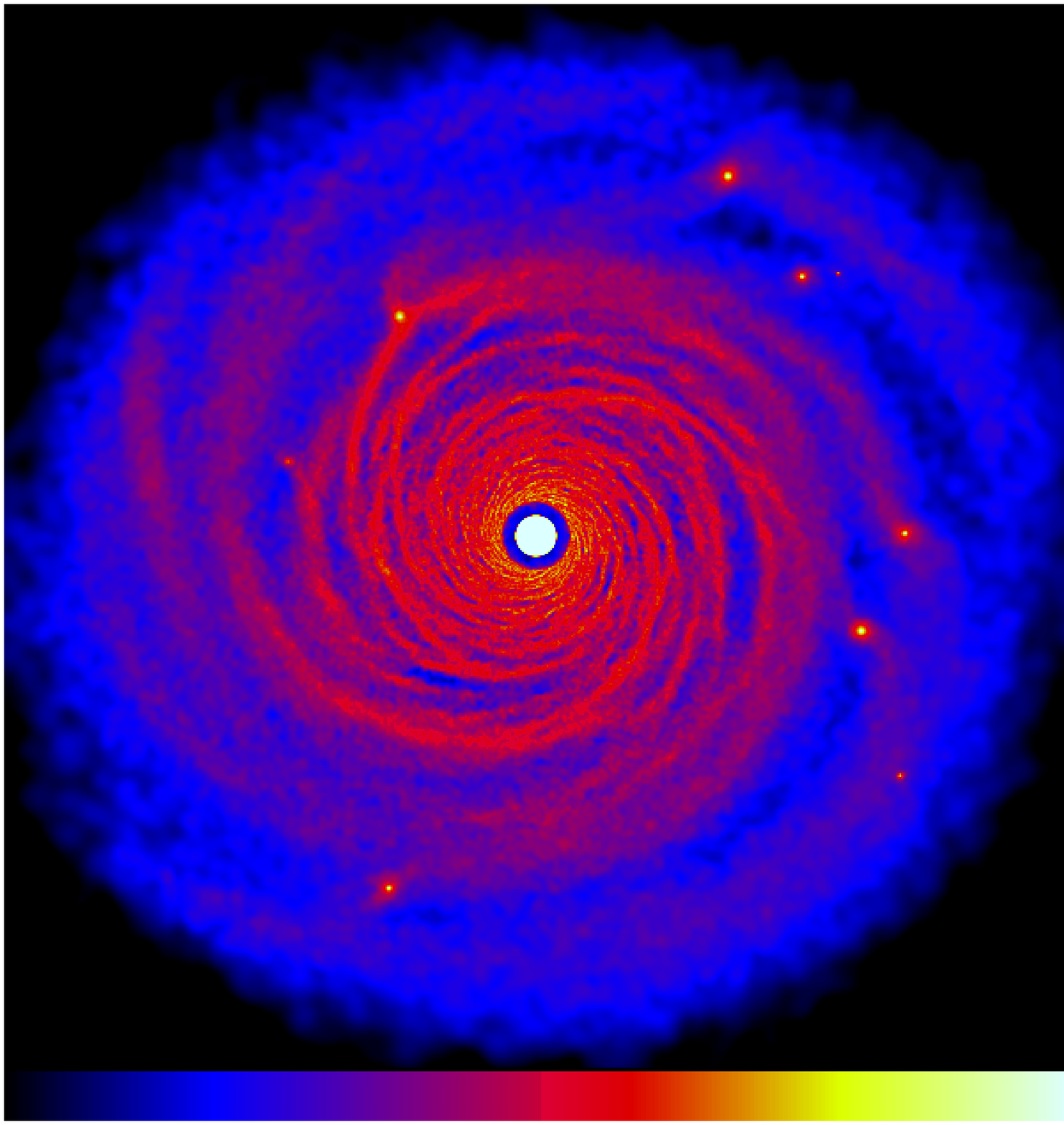,width=0.3\textwidth}}
\caption{\small { Images of Sh at fragmentation (left) and S1 (centre)
at the same  time. Although Sh has fragmented,  only one  fragment is seen at large radius and aside from the immediate area around the fragment the discs are very similar.
This should be contrasted with the profusion of fragments in F2 at the point
of fragmentation}}
 \label{fig:}
\end{figure*}

\subsection{Resolution issues}

One important aspect that needs to be taken into account is whether the resolution of our simulation is high enough to reproduce fragmentation, when it occurs. Resolution criteria for fragmentation with SPH codes have been discussed by \citet{bate97}. They obtained that SPH correctly reproduces fragmentation if the relevant Jeans mass contains at least 100 SPH particles, that is twice the typical number of neighbours ($N_{\rm neigh}=50$) within one smoothing region. More recently, \citet{nelson06} has revisited this issue focussing on fragmentation in self-gravitating discs and has found a slightly more stringent criterion, requiring that the Jeans mass is resolved with three times as many particles as required by \citet{bate97}. In a gravitationally unstable disc, the most unstable wavelength is given by $\lambda=2c_{\rm s}^2/G\Sigma$. The Jeans mass (or, as \citealt{nelson06} calls it, the ``Toomre mass'') is then given by:
\begin{equation}
M_{J}=\pi\Sigma\lambda^2= \frac{4\pi c_s^4}{G^2\Sigma} = 4\pi^3Q^2\left(\frac{H}{R}\right)^2
\Sigma R^2.
\end{equation}
The cumulative disc mass at radius $R$ is given by:
\begin{equation}
M_{\rm disc}(R)=2\pi\Sigma R^2=M_{\rm
disc}\frac{R}{R_{\rm out}},
\label{eq:discmass}
\end{equation}
since in our setup $\Sigma\propto R^{-1}$ (see above). We can then rewrite the Jeans mass using eq. (\ref{eq:discmass}), as:
\begin{equation}
M_{J}=
{2\pi^2}Q^2\left(\frac{H}{R}\right)^2\left(\frac{R}{R_{\rm out}}
\right)m_{\rm p}N_{\rm tot},
\end{equation}
where we have also used $M_{\rm disc}=m_{\rm p}N_{\rm tot}$, where $N_{\rm tot}$ is
the total number of particles used and $m_p$ is the mass of an
individual SPH particle. In order to properly resolve fragmentation, we
require that $M_T>m_pN_{\rm reso}$, where $N_{\rm reso}=2N_{\rm
neigh}=100$, according to \citet{bate97}, or $N_{\rm reso}=6N_{\rm
neigh}=300$ according to Nelson's more restrictive criterion, and
recalling that in our simulations the mean number of neighbours per
particle is 50. We then obtain that we have enough resolution at radii
$R$ that satisfy:
\begin{equation}
\frac{R}{R_{\rm out}}\gtrsim
\left(\frac{2}{\pi^2}\frac{1}{Q^4q^2}\frac{N_{\rm reso}}{N_{\rm tot}}
\right)^{1/3}, 
\label{eq:resolution}
\end{equation}
where $q=M_{\rm disc}/M_{\star}$ and where we have aso used the relationship between disc thickness and the parameter $Q$:
\begin{equation}
\frac{H}{R}=\frac{Q}{2} \frac{M_{\rm disc}(R)}{M_{\star}},
\label{eq:h}
\end{equation}
that can be easilty derived from Eq. (\ref{eq:Q}). 
Based on equation (\ref{eq:resolution}) we can then conclude that for $q=0.1$, as used in the present paper, fragmentation is well resolved at radii $R\gtrsim 5$. Note that, since $N_{\rm reso}$ only enters eq. (\ref{eq:resolution}) to the power of one third, if we had used the more restrictive condition of \citet{nelson06}, we would only increase our minimum radius by a factor 1.4. As shown in Fig. 2, whenever we observed fragmentation, this occurred outside $R\approx 5$, so that we can be confident that we do resolve the relevant mass and length scales for fragmentation.

A second aspect related to resolution is that we require artificial
viscosity to play a role only when modeling shocks. In order to ensure
this, we then require that the velocity difference accross a smoothing
kernel is subsonic, i.e. $h\Omega<c_{\rm s}$, where $h$ is the
smoothing length. This in turn requires that the smoothing length is
smaller than the disc thickness $H=c_{\rm s}/\Omega$. We have indeed
checked that, even at the lower resolution of 250,000 particles, the
average smoothing length is a fraction $\approx 0.5$ of the disc
thickness.

\section{Results}

  We find that in the fast case ($x=10.5$),  a fragment forms when
$\beta_{\rm cool}
= 0.75$, i.e. about $8.9$ outer disc dynamical times after the
rapid reduction in $\beta_{\rm cool}$ commenced.
Since fragmentation always takes about a dynamical timescale to
get under way, it follows that,
as expected, the `fast' case behaves like the usual
case where  a fixed $\beta_{\rm cool}$ is imposed.

  We however see different behaviour in the slow case: here
we find that when  $\beta_{\rm cool}$ is reduced to, and then
held at, $\beta _{\rm hold} = 2.75$ (the evolution of this simulation is not shown in Fig. 1), 
the disc does not fragment even when
the disc is then integrated for a further $44$ outer dynamical timescales.
Likewise, for the slow case, the disc does not fragment when held
at $\beta _{\rm hold} = 3$, even after integration for $63$
outer dynamical timescales at this $\beta _{\rm cool}$ value.
\footnote{We have tested whether this resistance to fragmentation
in the slow case is simply because the disc takes longer
to reach $\beta_{\rm cool} =3$ and is therefore of lower mass, due to
accretion onto the central star. However, in the  control run V (in which
the disc attains $\beta_{\rm cool} =3$ at the same time, but with
rapid ($x=10.5$) reduction in  $\beta_{\rm cool}$ between
$\beta_{\rm cool}=6$ and $\beta_{\rm cool}=3$), the disc fragments
promptly. Thus we are satisfied that it is indeed the value of
$x$ which controls fragmentation.} On the other hand, when $\beta_{\rm cool}$ 
was instead held at
$2.62$, it fragmented after a further $\sim 18$ outer dynamical timescales,
so it would appear that the fragmentation boundary is at around
$2.7$. This is  in strong contrast with the value of $\sim 7$ derived
in previous work  where a fixed $\beta_{\rm cool}$ is imposed. We
hesitate to say that we have proved that a disc will {\it never} fragment when
brought to such a low value of  $\beta_{\rm cool}$ value at this  slow rate,
since our experience shows that where one is close to the limit of 
marginally stable $\beta_{\rm cool}$, fragmentation
may ensue after long timescales, and that its timing may depend on
numerical noise that can be affected by  resolution. Indeed, we found that 
when we re-ran the $x=105$ simulation at higher resolution
($N=500,000$), and held it at $\beta_{\rm hold} = 3$,  it eventually did form
a fragment at large radius . We however show the  disc structure in this
simulation at the point of fragmentation and contrast it with the corresponding
situation when the cooling time is rapidly reduced and then held at
constant  $\beta_{\rm cool}=3$ (i.e. model F2). 
 Evidently, notwithstanding the fact that a fragment does
eventually form
in the former case also, the disc structure is quite different in the
two cases, with
the `rapid' simulation containing a number of regions that are on
the point of fragmentation at the moment that the first fragment
appears.  Our interest here is
not in defining precise boundaries at which fragmentation will or will
not occur
(since the definition of such a boundary is always contingent on the
duration of the simulation) but in demonstrating that the structure of
the
disc is indeed affected not just by the instantaneous value of
$\beta_{\rm cool}$ (and hence on the heating rate that has to
be delivered through the action of the self-gravitating modes) but also
on the {\it history} of how the disc arrived at such a value of
$\beta_{\rm cool}$.

This then raises the possibility (which we discussed in Section 1) that
the lower limit on $\beta_{\rm cool}$ for self-regulation might
represent
the difficulties that a disc might have in achieving a self-regulated
state on an appropriately short timescale, rather than a fundamental
upper limit on the dissipation rate that can be provided by
gravitational
modes in the absence of fragmentation. In principle, then,
we could envisage a situation where the disc might be self-regulated
at an arbitrarily low value of $\beta_{\rm cool}$ (i.e. where the
non-linear development of the spiral modes delivered an arbitrarily
high heating rate without the disc fragmenting) provided that the
disc approached this state sufficiently slowly. Such a conclusion
would contradict that of Lodato and Rice 2005, who interpreted
 the fragmentation boundary in terms of a (history independent)
limit on the maximum $\alpha $ value delivered by such instabilities.

  In order to explore this further, we ran the very slow ($x=314$) simulations
so that we could test whether the disc could remain self-regulated at
a yet smaller value of $\beta_{\rm cool}$ than for the $x=105$ case.
We however found little difference in the results for the $x=105$ and
$x=314$ case, the lowest values at which $\beta_{\rm cool}$ could
be held being respectively $2.75$ and $3.00$ for the two cases 
(for $N= 250,000$ in both cases).

\section{Conclusions}

We have found that the {\it rate} at which the cooling timescale is
changed indeed affects the minimum value of $\beta_{\rm cool}$
at which the disc can exist in a stable, self-regulated state. As
expected, this effect is only manifest when the the cooling
timescale is varied on a timescale ($\tau$)  that is longer than the
cooling timescale,
since for $\tau < t_{\rm cool}$, the temperature always falls
on a timescale $t_{\rm cool}$, irrespective of $\tau$. We find that
when $\tau > t_{\rm cool}$, the self-regulated state is sustainable at
cooling times that are about a factor two less than those that are
possible when a fixed cooling timescale is imposed at the outset of the
simulation. This implies that (in the slow cooling case) the
gravitational instabilities
are able to deliver about twice the heating rate without the disc
fragmenting.  In terms of the `viscous alpha' description
of such instabilities (Shakura and Sunyaev 1973, Gammie 2001,
Lodato \& Rice 2005), the maximum $\alpha$ deliverable by such a
disc is then increased from $\sim 0.06$ to $\sim 0.12$. It should be noted that such 
`local' description of the transport induced by gravitational instabilities
is only possible in the limit in which global, wave-like transport does not
play an important role. \citet{LR04,LR05}, using a cooling prescription similar 
to ours,  have shown that this is the case, as long as the total disc mass is 
small ($\lesssim 0.2M_{\star}$), which is the case for our simulations. 
\citet{mejia05}, using a constant cooling time, argue 
that global effects might be present, but do not explicitly calculate such global 
torques. On the other hand, recent calculation by \citet{boley06} (see in particular 
their Fig. 13), which employ more realistic cooling properties, confirm that in 
the limit of small disc mass, the transport 
induced by gravitational instabilities is essentially local. 

  We have thus found that thermal history can affect the ability of the
disc to exist in a self-regulated state without fragmentation but that
this affects the location of the stability boundary at only the factor
two level.
The fact that there was negligible change in the fragmentation
boundary when even slower changes in $t_{\rm cool}$ were
employed, demonstrates that thermal history is only part of the story.
Our results suggest that, however slowly the disc is cooled through a
sequence
of thermal equilibria, there is still a fundamental upper limit to
the heating that can be provided by gravitational instabilities
in a non-fragmenting disc. Thus it would appear that the initiation
of fragmentation in a self-gravitating disc does {\it not}
require that the disc enter the regime of rapid cooling on 
a {\it short} timescale. It is thus
 unnecessary to invoke sudden events (e.g. impulsive
interactions with passing stars, \citealt{LMC07})
to tip a previously self-regulated disc into the fragmenting regime.
Instead our results suggest  that fragmentation can in principle be approached
via
the {\it secular}  evolution of a self-gravitating disc.

\section*{Acknowledgements}

The simulations presented in this work have been performed at the UK
Astrophysical Fluid Facility (UKAFF). We thank Richard Durisen and Ken Rice
 for interesting discussions

\bibliographystyle{mn2e}

\bibliography{lodato}

\end{document}